\documentclass[sigconf,nonacm]{acmart}

\usepackage{tabularx}

\usepackage{multirow}
\usepackage{tikz}
\usepackage{amsmath}
\usepackage{wrapfig}
\usepackage{float}
\usepackage{pifont}
\usepackage{colortbl}
\usepackage{booktabs, siunitx}
\usepackage{wasysym}
\usepackage{fontawesome}
\usepackage{color, colortbl}
\usepackage{array}
\usepackage{algorithmic}
\usepackage{graphicx}
\usepackage{textcomp}
\usepackage{xcolor}
\usepackage{pifont}
\usepackage{colortbl}
\usepackage{wasysym}

\usepackage{booktabs, siunitx}
\usepackage{color, colortbl}
\definecolor{Gray}{gray}{0.9}
\definecolor{LightCyan}{rgb}{0.88,1,1}
\usepackage{arydshln}
\usepackage{caption}
\usepackage{supertabular}
\usepackage{subcaption}
\usepackage{tcolorbox}
\usepackage{lipsum}

\usepackage{url}
\usepackage{tcolorbox}
\tcbuselibrary{skins, breakable}

\usepackage{listings} %
\usepackage{microtype} %
\usepackage{ragged2e} %

\renewcommand\footnotetextcopyrightpermission[1]{} %
\usepackage[linesnumbered,vlined]{algorithm2e}
\usepackage{graphicx} %
\usepackage{makecell}
\usepackage{enumitem}
\usepackage{subcaption}
\usepackage{changepage}

\newcolumntype{P}[1]{>{\centering\arraybackslash}m{#1}} 

\definecolor{nickgreen}{HTML}{007007}
\definecolor{dgreen}{HTML}{03A60D}
\definecolor{dred}{HTML}{A6030D}
\definecolor{darkbrown}{rgb}{1, 0.55, 0}

\definecolor{Gray}{gray}{0.9}
\definecolor{LightCyan}{rgb}{0.88,1,1}
\definecolor{Color001}{RGB}{148,0,85}
\definecolor{Color002}{RGB}{0,120,120}
\definecolor{Color003}{RGB}{199,232,172}

\AtBeginDocument{%
  }

\begin{document}

\title{SoK: Evolution, Security, and Fundamental Properties of Transactional Systems}

\author{Sky Pelletier Waterpeace}
\email{waterpeace@rowan.edu}
\orcid{0000-0002-2231-0160}
\affiliation{%
  \institution{Rowan University}
  \city{Glassboro}
  \state{New Jersey}
  \country{USA}
}

\author{Nikolay Ivanov}
\orcid{0000-0002-2325-2847}
\email{ivanov@rowan.edu}
\affiliation{%
  \institution{Rowan University}
  \city{Glassboro}
  \state{New Jersey}
  \country{USA}
}

\begin{abstract}
Transaction processing systems underpin modern commerce, finance, and critical infrastructure, yet their security has never been studied holistically across the full evolutionary arc of these systems. Over five decades, transaction processing has progressed through four distinct generations, from centralized databases, to distributed databases, to blockchain and distributed ledger technologies (DLTs), and most recently to multi-context systems that span cyber-physical components under real-time constraints. Each generation has introduced new transaction types and, with them, new classes of vulnerabilities; successful exploits now cause billions of dollars in annual losses. Despite this, security research remains fragmented by domain, and the foundational ACID transaction model has not been revisited to reflect the demands of contemporary systems.

In this Systematization of Knowledge, we survey 235 papers on transaction security, classify 163 in-scope works by evolutionary generation, security focus (attack, defense, or vulnerability), and relevant Common Weakness Enumeration (CWE) entries, and distill a curated set of 41 high-impact or seminal papers spanning all four generations. We make three principal contributions. First, we develop a four-generation evolutionary taxonomy that contextualizes each work within the broader trajectory of transaction processing. Second, we map each paper's security focus to CWE identifiers, providing a systems-oriented vocabulary for analyzing transaction-specific threats across otherwise siloed domains. Third, we demonstrate that the classical ACID properties are insufficient for modern transactional systems and introduce \emph{RANCID}, extending ACID with \emph{Real-timeness}~(R) and \emph{$N$-many Contexts}~(N), as a property set for reasoning about the security and correctness of systems that must coordinate across heterogeneous contexts under timing constraints. Our systematization exposes a pronounced bias toward DLT security research at the expense of broader transactional security and identifies concrete open problems for the next generation of transaction processing systems.
\end{abstract}

\keywords{Cyber Systems, Computer Systems, Transaction Processing Systems}

\maketitle

\section{Introduction}\label{sec:introduction}

Transaction processing is foundational to modern digital infrastructure. From banking and e-commerce to supply-chain logistics and decentralized finance, virtually every domain that manages shared mutable state relies on some form of transaction processing~\cite{grayTransactionConceptVirtues1981, grayTransactionProcessingConcepts1994, kleppmannDesigningDataintensiveApplications2017}. What appears to an end user as a single operation, such as purchasing an item from an online store, is in practice a complex web of interdependent subtransactions spanning the customer's bank, a payment processor, the merchant's inventory system, one or more shipping providers, and notification services. Each of these subsystems may itself be distributed, replicated, or decentralized, and a failure or compromise in any one of them can cascade across the others.

Despite this centrality, the security of transaction processing systems has never been studied \textit{holistically}. Individual research communities have addressed security within their respective domains: the database community has investigated concurrency-control attacks and access-control models~\cite{abrahamTransactionSecuritySystem1991, jajodiaMultilevelSecureTransaction2001, ammannRecoveryMaliciousTransactions2002}, the payment-systems community has exposed protocol-level flaws in EMV and contactless payment schemes~\cite{murdochChipPINBroken2010, bondChipSkimCloning2014, basinEMVStandardBreak2021}, and the blockchain community has cataloged a wide range of vulnerabilities in consensus protocols, smart contracts, and decentralized exchanges~\cite{atzeiSurveyAttacksEthereum2017, saadExploringAttackSurface2020, chenSurveyEthereumSystems2020, qinAttackingDeFiEcosystem2021}. Existing surveys reflect this siloing: they typically focus on a single domain such as smart-contract vulnerabilities~\cite{atzeiSurveyAttacksEthereum2017, chenSurveyEthereumSystems2020}, DeFi-specific attacks~\cite{qinAttackingDeFiEcosystem2021}, or blockchain consensus~\cite{saadExploringAttackSurface2020}, without connecting the security challenges that arise from the \emph{transactional nature} common to all these systems. No unifying framework bridges these communities.

This gap is consequential. Successful attacks on transactional systems cause billions of dollars in losses annually, over \$2.2B~USD in cryptocurrency theft alone in 2024~\cite{trm2025CryptoCrime}, in addition to large-scale fraud enabled by weaknesses in payment-card transaction protocols~\cite{bondAttacksCryptoprocessorTransaction2001, murdochChipPINBroken2010, eurpeanbankingauthority2025ReportPayment2025}. As transactional systems continue to grow in complexity and interconnectedness, understanding the shared structure of their security challenges becomes increasingly urgent.

A principal reason for this fragmentation is that transactional systems have undergone rapid and fundamental evolution. Early transaction processing systems were centralized databases serving multiple users within a single organization~\cite{astrahanSystemRelationalApproach1976, eswaranNotionsConsistencyPredicate1976}. The introduction of distributed databases added inter-database coordination and replication~\cite{bernsteinConcurrencyControlDistributed1981, bernsteinConcurrencyControlRecovery1987, thomasMajorityConsensusApproach1979}. Later, Distributed Ledger Technologies (DLTs) such as blockchain introduced nonfederated consensus, removing the assumption of a central authority entirely~\cite{garayBitcoinBackboneProtocol2024, woodETHEREUMSECUREDECENTRALISED, androulakiHyperledgerFabricDistributed2018}. Most recently, modern transactional systems frequently span multiple heterogeneous contexts, including cyber-physical components and real-time constraints, as seen in autonomous vehicle control, industrial IoT, and cross-platform DeFi protocols~\cite{duoSurveyCyberAttacks2022, daianFlashBoys202020, rahmadikaSecurityAnalysisDecentralized2018}. Each evolutionary stage has introduced new categories of transactions and, with them, new classes of vulnerabilities. To make sense of this progression, we develop a taxonomy of four evolutionary generations of transaction processing systems and classify the security literature accordingly.

Compounding the challenge, no existing security framework is designed to apply across all forms of transaction processing. Standards such as the Smart Contract Security Verification Standard (SCSVS) and EEA EthTrust Certification target smart contracts specifically, while the OWASP Application Security Verification Standard (ASVS) addresses web applications. To bridge this gap, we treat transaction processing as an architectural layer in system design and adopt the Common Weakness Enumeration (CWE) system~\cite{CWEEvolutionCWE}, a comprehensive and continuously maintained catalog of software and hardware weaknesses, as a principled basis for classifying the security focus of each paper in our survey. This approach situates transaction security within the broader discipline of systems security and provides a common vocabulary for comparing vulnerabilities across otherwise disparate transactional domains.

Our cross-generational analysis further reveals a limitation in the foundational model of transaction properties. The ACID properties (Atomicity, Consistency, Isolation, and Durability) codified during the era of centralized databases~\cite{grayTransactionConceptVirtues1981}, have served as the backbone of transaction processing theory for over four decades~\cite{grayTransactionProcessingConcepts1994, abrahamTransactionSecuritySystem1991, netoSecurityBenchmarkingTransactional2012}. However, ACID was designed for an era in which transactions operated within a single system context and without hard timing constraints. Modern transactional systems routinely violate both assumptions: an autonomous drone must execute state transitions across multiple control subsystems within strict deadlines, and a cross-chain DeFi swap must coordinate among independent ledgers under time-sensitive arbitrage conditions~\cite{zhangEfficientDistributedTransaction2023, daianFlashBoys202020}. We therefore extend ACID with two additional properties, Real-timeness~(R) and $N$-many Contexts~(N), yielding the RANCID framework, and we assess which RANCID properties are affected by or related to the security focus of each surveyed work.

\begin{figure}
    \centering
    \includegraphics[width=1\linewidth]{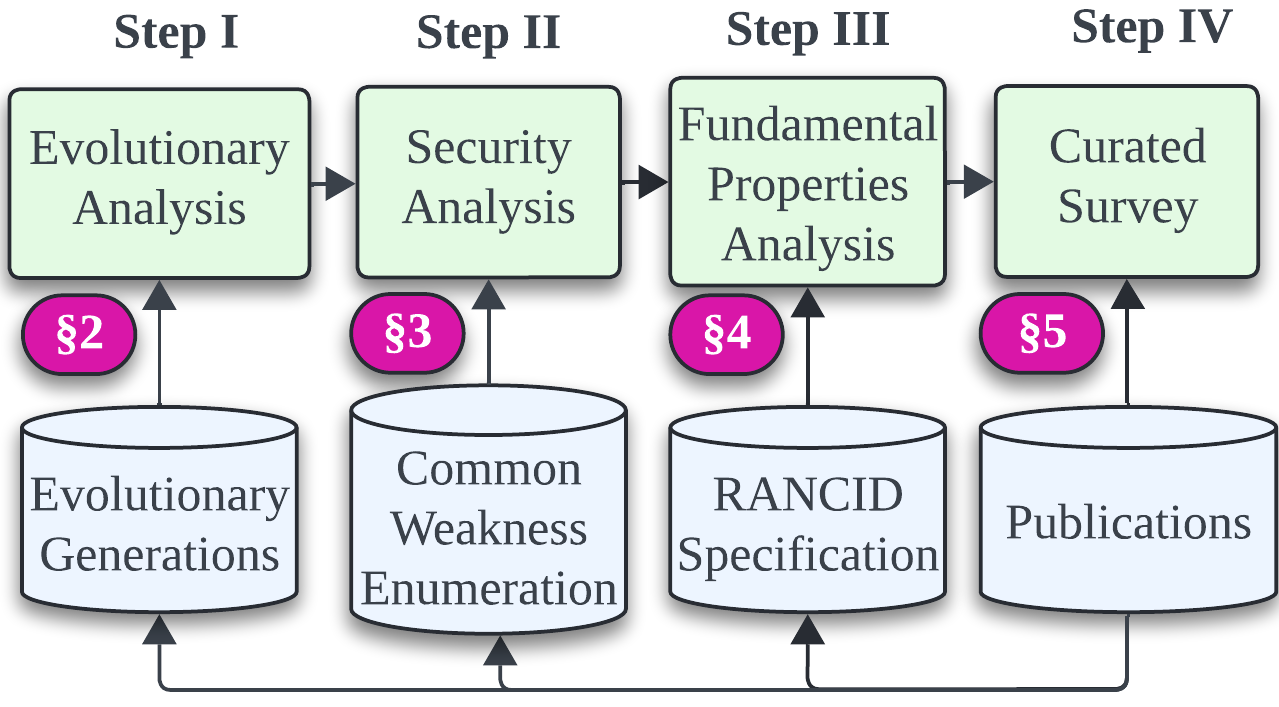}
    \caption{Overview of our systematization approach. Each step corresponds to a major section of the paper and produces a reusable analytical artifact (bottom row).}
    \label{fig:overview}
\end{figure}

An overview of our approach is illustrated in Figure~\ref{fig:overview}. In summary, this work makes the following contributions:
            \begin{itemize}[leftmargin=*]
                \item \textbf{Evolutionary taxonomy (\S\ref{sec:evo}).} We identify four generations of transaction processing systems, from centralized databases to modern multi-context systems, and classify the transaction security literature by generation, exposing a pronounced bias toward DLT research.
                \item \textbf{CWE-based security classification (\S\ref{sec:sec}).} We map each surveyed paper to Common Weakness Enumeration entries, providing a systems-oriented vocabulary for analyzing transaction-specific vulnerabilities, attacks, and defenses across domains.
                \item \textbf{RANCID properties (\S\ref{sec:rancid}).} We demonstrate that ACID is insufficient for modern transactional systems and introduce RANCID, extending ACID with Real-timeness and $N$-many Contexts, as a property set for reasoning about security and correctness under contemporary demands.
                \item \textbf{Curated survey (\S\ref{sec:classification}).} We distill 41 high-impact or seminal papers spanning five decades into a classified reference organized along all dimensions of our taxonomy, providing a structured entry point for researchers new to this area.
            \end{itemize}

\section{Evolutionary Generations of Transaction Processing Systems}\label{sec:evo}

To provide a unified lens for analyzing security across diverse transactional domains, we identify four major generations in the evolution of transaction processing systems. Each generation introduces new categories of transactions, new trust assumptions, and correspondingly new security considerations. Table~\ref{tab:generations} summarizes the four generations, which we describe in detail below.

\paragraph{Generation I: Centralized Databases.}
The earliest transaction processing systems were designed to allow multiple users to interact concurrently with a shared data store~\cite{astrahanSystemRelationalApproach1976, eswaranNotionsConsistencyPredicate1976}. A canonical example is a bank maintaining a single database of customer accounts, accessible to employees through multiple terminals. The central challenge in such systems is ensuring that concurrent transactions do not interfere with one another: each user must be able to access the database as though no other users are present, updates must either complete in their entirety or have no effect, all changes must respect the business logic of the system, and committed changes must survive system failures. These requirements are formalized as the ACID properties: Atomicity, Consistency, Isolation, and Durability~\cite{grayTransactionConceptVirtues1981, grayTransactionProcessingConcepts1994}.

From a security perspective, Generation~I systems operate under a trust model in which the database administrator and the hosting organization are implicitly trusted. The principal security threats are from insider abuse, where an attacker who can submit transactions to the shared store may exploit race conditions or bypass access controls to read or modify data they should not be able to reach~\cite{abrahamTransactionSecuritySystem1991}. Because all transactions flow through a single system, the attack surface is comparatively narrow, but the concentration of authority makes the consequences of a compromise severe.

\paragraph{Generation II: Distributed Databases.}
As organizations grew, the need arose for the data store itself to be replicated across multiple systems, often in geographically distinct locations, in order to provide fault tolerance and reduce access latency for remote users~\cite{bambergerCiticorpsNewHighperformance1989, bernsteinConcurrencyControlRecovery1987, clarkComparisonCommercialMilitary1987}. In Generation~II the original challenge of concurrent multi-user access is compounded by the requirement that multiple copies of the database remain synchronized~\cite{thomasMajorityConsensusApproach1979, bernsteinConcurrencyControlDistributed1981}.

This synchronization requirement gives rise to new classes of transactions. In addition to user-to-database operations, distributed systems require database-to-database transactions (usually operating automatically behind the scenes) to maintain a coherent shared state across replicas~\cite{mossNESTEDTRANSACTIONSAPPROACH1981, bernsteinFormalAspectsSerializability1979, reedNamingSynchronizationDecentralized1978}. These inter-replica transactions introduce complications distinct from user-facing ones. For example, if one or more replicas lose network connectivity, the system must determine the authoritative values for records that may have been updated independently on partitioned nodes, a problem that requires notions of distributed consensus~\cite{breitwieserDistributedTransactionProcessing1982}.

The classes of database users also diversify in Generation~II. Returning to the bank example, customers now interact with the system directly through ATMs, telephone banking, and early online interfaces, in addition to the employee-mediated access of Generation~I. The ACID properties continue to apply but must now be enforced across distributed infrastructure, covering not only multi-class user-database interactions but also the newly introduced inter-database transactions.

The move to distributed architectures substantially expands the attack surface. Inter-replica communication channels become targets for eavesdropping and message injection; network partitions can be exploited to cause inconsistent state across replicas; and the introduction of multiple administrative domains raises questions about trust boundaries and access-control enforcement that were absent in the single-system setting~\cite{clarkComparisonCommercialMilitary1987, gilbertBrewersConjectureFeasibility2002}. The Clark--Wilson integrity model~\cite{clarkComparisonCommercialMilitary1987}, for instance, arose from the need to formalize transaction integrity guarantees in commercial systems where multiple parties interact with shared data across organizational boundaries.

\paragraph{Generation III: Distributed Ledger Technologies.}
As computer hardware became more affordable, more businesses adopted database technology~\cite{diasCouplingManySmall1986}, and concerns about privacy and centralized control of personal information intensified. Already in the early years of the distributed database era, researchers had begun exploring ways to remove centralized authority from data management~\cite{chaumSecurityIdentificationTransaction1985a}. Although it took several decades, these efforts ultimately led to the development of blockchain, a method of consensus-based shared data management without any central authority, and the first major cryptocurrency, Bitcoin~\cite{nakamotoBitcoinPeertoPeerElectronic2008}.

The introduction of Bitcoin sparked an explosion of related technologies: Ethereum, which enables automatic execution of state transitions through smart contracts~\cite{woodETHEREUMSECUREDECENTRALISED}; Hyperledger Fabric, a modular, permissioned variant designed for enterprise applications~\cite{androulakiHyperledgerFabricDistributed2018}; Monero, a privacy-focused cryptocurrency~\cite{kumarTraceabilityAnalysisMoneros2017}; and many further extensions~\cite{liSurveyStateoftheartSharding2023, hedayatiSurveyBlockchainChallenges2021}. These systems are known collectively as Distributed Ledger Technologies (DLTs)~\cite{uzomaComprehensiveReviewMultiCloud2025} and are characteristic of Generation~III.

In the DLT era, the data store is maintained collectively by a consortium of participating \emph{nodes} rather than by a single organization. The ledger may be fully public, as with Bitcoin and Ethereum, or permissioned, as in supply-chain deployments among a closed group of companies~\cite{kaurResearchSurveyApplications2021}. Compared to earlier generations, DLTs rely on a fundamentally new type of transaction: decentralized consensus, in which participating nodes must agree on whether submitted transactions are valid and on the authoritative state of the ledger, without any single trusted party~\cite{bouragaTaxonomyBlockchainConsensus2021}.

The introduction of smart contracts further expands the transaction model by enabling programmatic, self-executing logic embedded in the ledger~\cite{buterinEthereumNextGenerationSmart}. Ethereum, for example, provides a Turing-complete smart contract language~\cite{hildenbrandtKEVMCompleteSemantics2017}, meaning that in principle any state transition can be expressed as a smart contract. Smart contracts have been leveraged for flash loans, funds borrowed and repaid within a single block time~\cite{daianFlashBoys202020}, and Decentralized Autonomous Organizations (DAOs), corporation-like entities governed entirely by smart contract code~\cite{shierUnderstandingRevolutionaryFlawed2017}. However, smart contracts have also been extensively exploited for attacks on other contracts, causing losses in the billions of USD~\cite{abdelazizSchoolingExploitFoolish2023, aftabSecurityAnalysisOnline2024, trm2025CryptoCrime}. Generation~III has therefore attracted a disproportionately high volume of security research (see Figure~\ref{fig:gens_bar}), driven by high-profile exploits and the substantial financial stakes involved~\cite{shierUnderstandingRevolutionaryFlawed2017, saadExploringAttackSurface2020}.

\paragraph{Generation IV: Modern Multi-Context Transactional Systems.}
Although distributed databases remain ubiquitous and companies continue to explore DLT integration~\cite{alqaisiTrustworthyDecentralizedLast2023, khanSecurityAnalysisBlockchain2023, rahmadikaSecurityAnalysisDecentralized2018}, many modern transactional systems have evolved beyond these modalities. The most recent generation is characterized by the need to coordinate transactions across multiple heterogeneous contexts, often under real-time constraints. Consider a DeFi protocol that orchestrates a flash-loan transaction spanning multiple independent smart contracts, liquidity pools, and price oracles, all within a single atomic block~\cite{daianFlashBoys202020, chenFlashSynFlashLoan2024}. Or consider an autonomous vehicle whose navigation depends on state transitions across sensor subsystems, control actuators, and communication modules, each operating on distinct timescales and requiring coordinated transactional guarantees~\cite{duoSurveyCyberAttacks2022}. Modern transactional systems also include cross-platform payment ecosystems in which a single consumer purchase involves coordinated transactions among banks, card networks, merchants, and logistics providers~\cite{markantonakisUsingAmbientSensors2024, basinEMVStandardBreak2021}, as well as industrial control systems in which transactional integrity has direct physical consequences~\cite{saidCyberAttackP2PEnergy2022, liangGeneralizedFDIABasedCyber2018}.

A common feature that unifies these diverse systems is the large number of distinct contexts across which a single logical transaction must be processed. Distributed databases of Generation~II added database-to-database transactions, and DLTs of Generation~III added node-to-node consensus and smart-contract interactions. However, Generation~IV systems are fundamentally multicontextual: a given transaction may require coordination not only among multiple computer systems and databases, but also physical sensors, relays, industrial controllers, GPS satellites, and more. Modern transactional systems intrinsically require transaction processing across $N$-many distinct contexts.

The security implications of this multicontextual nature are profound. Each additional context introduces its own trust assumptions, failure modes, and communication channels, all of which become potential attack vectors. A compromise in any single context, such as a spoofed sensor reading, a manipulated price oracle, or a delayed control signal, can propagate through the transactional chain and undermine the integrity of the overall operation~\cite{duoSurveyCyberAttacks2022, liangGeneralizedFDIABasedCyber2018}. Moreover, enforcing consistent access-control policies across heterogeneous contexts with different administrative domains and security models remains an open problem with direct relevance to the access-control community.

A second distinguishing feature of Generation~IV systems is the frequent reliance on firm timing constraints. Whereas earlier generations occasionally had or aspired to real-time properties~\cite{lelannDistributedSystemRealTime1981, stankovicRealtimeTransactions1988, eTimedrivenSchedulerRealtime1986}, such constraints were often driven by usability concerns, with only specialized applications requiring hard real-time transaction processing~\cite{laplanteHistoricalSurveyEarly1995}. By contrast, hard real-time constraints are now commonplace: autonomous vehicles must complete control-loop transactions within milliseconds, and services such as Uber continuously process petabytes of data in real time across heterogeneous subsystems~\cite{fuRealtimeDataInfrastructure2021, buttersteinReplicationSpeedChange2020, caoTitAntOnlineRealtime2019}. The need to process transactions across multiple distinct contexts while satisfying real-time constraints is not captured by the classical ACID model and motivates the extended RANCID framework introduced in \S\ref{sec:rancid}.

\begin{tcolorbox}[title=\textbf{Insight from Evolutionary Analysis}, colback=white!95!black, colframe=gray!75!black, colbacktitle=gray!75!black, coltitle=white, boxsep=1.5pt, left=1mm, right=1mm, top=1mm, bottom=1mm]
Transaction processing systems have evolved through four generations, each introducing new categories of transactions and expanding the attack surface. The disproportionate research focus on Generation~III (DLTs) has left fundamental challenges in modern multi-context transactional systems underexplored.
\end{tcolorbox}

\begin{table}
    \centering
        \caption{Generations of Transaction Processing Systems}

    \begin{tabular}{m{.3cm}| >{\RaggedRight\arraybackslash}m{1.9cm} :>{\RaggedRight\arraybackslash}m{2.4cm}: >{\RaggedRight\arraybackslash}m{2.3cm} }
            \multicolumn{1}{c}{\textbf{Era}} & \multicolumn{1}{c}{\textbf{System Type}} & \multicolumn{1}{c}{\textbf{Key Features}} & \multicolumn{1}{c}{\textbf{New Txn Types}}
            \\
            \hline
                                
          \multicolumn{1}{c|}{\Large \textbf{I}} 
                     & Centralized Databases & Concurrent multi-user access to shared data & User--database
                     \\
                     \hdashline
          \multicolumn{1}{c|}{\large \textbf{II}}  
                    & Distributed Databases 
                    & Data replicated across locations and organizations & Inter-database (replication, sync), multiple user classes
                     \\
                     \hdashline
         \multicolumn{1}{c|}{\large \textbf{III}}  
                    & Distributed Ledgers & Nonfederated data storage relying on consensus protocols & Node--node (authentication and consensus), smart contracts
                     \\
                     \hdashline
         \multicolumn{1}{c|}{\large \textbf{IV}} 
                    & Modern Multi-Context TPS & Transactions span heterogeneous cyber-physical systems with real-time constraints & Cross-context ($N$-many), real-time
                     \\
        \hline
    \end{tabular}
    \label{tab:generations}
\end{table}

\section{Security of Transaction Systems}\label{sec:sec}

Security of transactional systems is of critical importance given the ubiquity of these systems in modern commerce and infrastructure. High-profile attacks on transactional systems have resulted in billions of dollars in cumulative losses~\cite{agrafiotisTaxonomyCyberharmsDefining2018, trm2025CryptoCrime, eurpeanbankingauthority2025ReportPayment2025}, and the attack surface continues to expand as systems grow in complexity. In this section we survey the landscape of transaction security threats across evolutionary generations and then present our CWE-based classification framework for analyzing these threats systematically.

\subsection{Landscape of Transaction Security Threats}\label{sec:sec:landscape}

\paragraph{Payment system security.}
A long-standing area of concern is the security of credit card payments. The EMV standard, used in billions of cards worldwide, was introduced to secure in-person payments through a combination of a hardware security chip and a cardholder PIN~\cite{basinEMVStandardBreak2021}. Despite the enormous effort invested in the protocol\footnote{The specification exceeds 2,000 pages.} numerous flaws have been discovered that allow attackers to bypass its intended guarantees. These include relay attacks that enable a purchaser to use a victim's card for an in-person purchase anywhere in the world~\cite{bocekNFCRelayAttack2016, markantonakisUsingAmbientSensors2024, gurulianPreventingRelayAttacks2017}, logical flaws in the protocol that permit authentication bypass~\cite{basinCardBrandMixup2021, murdochChipPINBroken2010, bondChipSkimCloning2014}, and hardware attacks on the random number generators used in card readers~\cite{markettosFrequencyInjectionAttack2009}. Total worldwide losses from payment fraud are estimated at multiple billions of dollars annually~\cite{eurpeanbankingauthority2025ReportPayment2025}.

\paragraph{Distributed ledger security.}
The introduction of DLTs created both novel applications and corresponding avenues for theft and fraud. The 2016 DAO attack exemplifies both: The DAO was a venture capital fund governed entirely by smart contracts, embodying the early Ethereum community's philosophy that ``Code is Law''~\cite{shierUnderstandingRevolutionaryFlawed2017}. Within two months it had secured \$168M~USD in investments, but an attacker exploited a reentrancy flaw to drain over \$50M~USD, ultimately precipitating a hard fork that split Ethereum into two chains. The DAO attack was an early signal of a much larger trend: over \$2.2B~USD was stolen from crypto ecosystems in 2024 alone~\cite{trm2025CryptoCrime}.

\begin{figure}
    \centering
    \includegraphics[width=0.95\linewidth]{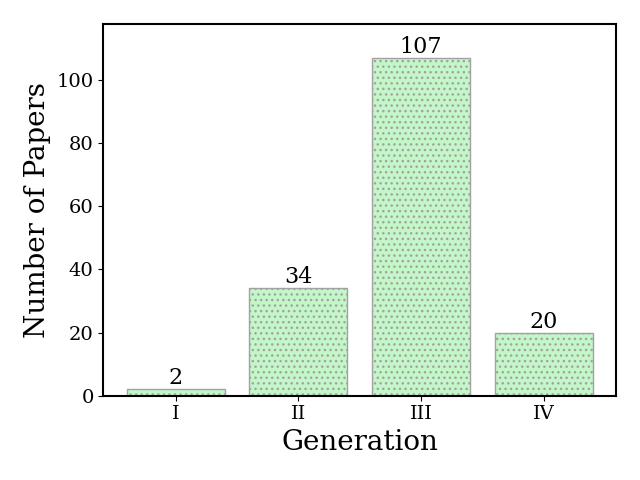}
    \caption{Distribution of security-focused papers across evolutionary generations in our survey ($n = 163$). Generation~III (DLTs) accounts for 66\% of the literature.}
    \label{fig:gens_bar}
    \vspace{-5pt}
\end{figure}

A distinctive feature of smart contract security is that contracts deployed on immutable ledgers cannot be patched, meaning that any vulnerability is permanent once the contract is live. This has given rise to an ecosystem of smart contract--mediated attacks~\cite{atzeiSurveyAttacksEthereum2017, wuBlockchainSecurityThreats2025}, automated vulnerability detection tools~\cite{tsankovSecurifyPracticalSecurity2018, chenSODAGenericOnline2020}, retroactive protection mechanisms for already-deployed contracts~\cite{rodlerSereumProtectingExisting2018}, and even automatic attack synthesis~\cite{fengPreciseAttackSynthesis2019, abdelazizSchoolingExploitFoolish2023}. As introduced in \S\ref{sec:evo}, flash loan attacks represent a further evolution, exploiting cross-contract coordination and tight timing constraints to create artificial arbitrage opportunities~\cite{daianFlashBoys202020, chenFlashSynFlashLoan2024, caoFlashotSnapshotFlash2021}. The enormous financial stakes have driven a disproportionate focus on Generation~III security research relative to other transactional domains, as shown in Figure~\ref{fig:gens_bar}.

\paragraph{Cyber-physical transaction security.}
Whereas earlier generations were exposed primarily to informational and financial threats, modern cyber-physical transactional systems face a broader threat landscape. Ukraine in 2015 experienced the first documented cyber attack against electricity providers, leaving over 200,000 people without power~\cite{robertm.leeAnalysisCyberAttack2016}, and threats to electric grids and other infrastructure have continued to grow~\cite{cyberthreatintelligenceintegrationcenterRecentCyberAttacks2024}. The proliferation of IoT devices provides additional attack surface~\cite{kumarDistributedIntrusionDetection2022}. Beyond denial of service, cyber-physical systems are vulnerable to transactional fraud, such as combined false data injection and topology attacks on electricity markets that subtly raise wholesale prices to defraud consumers~\cite{liangGeneralizedFDIABasedCyber2018}. These Generation~IV threats are particularly concerning because they can have direct physical consequences, extending the impact of transaction security failures beyond the digital domain.

\subsection{CWE-Based Classification Framework}\label{sec:sec:cwe}

Existing efforts to systematize transaction security remain domain-specific. The EEA EthTrust Certification~\cite{ethtrustsl} and   Smart Contract Security Verification Standard (SCSVS)~\cite{SmartContractSecurity2024} both target smart contracts, while the OWASP Application Security Verification Standard (ASVS)~\cite{OWASPApplicationSecurity} addresses web applications. No comprehensive framework exists for reasoning about transaction security across all system generations.

To bridge this gap, we treat transactions as an architectural layer in system design and adopt the Common Weakness Enumeration (CWE) system~\cite{CWEEvolutionCWE}, a comprehensive and continuously maintained catalog of software and hardware weaknesses, as the basis for classifying the security focus of each paper in our survey. For each of the 163 in-scope papers, we identified the principal security focus (attack, defense, or vulnerability) and the CWE entries most closely related to the security issues addressed. This approach situates transaction security within the broader discipline of systems security and provides a common vocabulary for comparing threats across otherwise disparate transactional domains.

\subsection{Patterns in the CWE Classification}\label{sec:sec:patterns}

Mapping the 41 curated papers to CWE entries (Table~\ref{tab:classification}) reveals several notable patterns. Across these works, we identified 61 unique CWE entries. Grouping them thematically exposes the dominant classes of weaknesses in transaction security.

\paragraph{Race conditions and timing weaknesses.}
The most prevalent weakness class involves race conditions and timing-related issues. CWE-362 (Race Condition) and CWE-367 (Time-of-Check Time-of-Use) appear in 9 and 6 curated papers, respectively, while CWE-385 (Covert Timing Channel) appears in 3. This dominance is unsurprising: transaction processing is fundamentally about coordinating concurrent access to shared state, and any failure in that coordination can be exploited. Race conditions underlie threats ranging from double-spending attacks on blockchains~\cite{karameTwoBitcoinsPrice2012} to front-running in decentralized exchanges~\cite{daianFlashBoys202020, zhangCombattingFrontRunningSmart2023} and write-skew anomalies in distributed databases. The prominence of timing-related CWEs also reinforces the motivation for including Real-timeness as a fundamental property in our RANCID framework (\S\ref{sec:rancid}).

\paragraph{Business logic and emergent-resource weaknesses.}
CWE-1229 (Creation of Emergent Resource) is tied for the most frequent, appearing in 9 papers, and CWE-841 (Improper Enforcement of Behavioral Workflow) appears in 5. Together with CWE-1039 (Automated Recognition Mechanism with Inadequate Detection), these weaknesses capture a class of threats in which system behavior diverges from designer intent not because of traditional implementation bugs, but because of unanticipated interactions between correctly functioning components. The DAO attack, flash loan exploits, and selfish mining strategies all fall into this category. These weaknesses are particularly difficult to address because they often arise only at the level of composed transactions spanning multiple subsystems, a pattern that becomes increasingly common in Generation~IV systems.

\paragraph{Authentication and authorization weaknesses.}
CWE-290 (Authentication Bypass by Spoofing) appears in 5 papers and CWE-294 (Authentication Bypass by Capture-Replay) in 4, reflecting persistent challenges in transaction authentication across domains. In payment systems, relay and replay attacks bypass EMV's intended authentication guarantees~\cite{murdochChipPINBroken2010, bondChipSkimCloning2014, markantonakisUsingAmbientSensors2024}. In DLTs, spoofing and Sybil-style attacks can subvert consensus~\cite{apostolakiHijackingBitcoinRouting2017}. CWE-284 (Improper Access Control) also recurs, underscoring that access-control enforcement %
remains a cross-cutting challenge in transactional systems, particularly as transactions span multiple administrative domains with heterogeneous authorization models.

\paragraph{Cross-generation observations.}
An important finding from the CWE mapping is that many weakness classes recur across generations, albeit manifesting differently. Race conditions appear in Generation~I concurrency-control exploits, Generation~II partition-based attacks, Generation~III front-running and double-spending, and Generation~IV cross-context timing attacks. This recurrence validates our decision to adopt CWE as a unifying vocabulary: it reveals structural commonalities in transaction security that domain-specific frameworks would obscure. At the same time, the growing prevalence of emergent-resource and behavioral-workflow weaknesses (CWE-1229, CWE-841) in later generations suggests that as systems grow more complex and multicontextual, the threat model shifts from implementation-level bugs toward composition-level failures, evidence of a trend that demands new analytical approaches.

\paragraph{Gaps and underrepresented areas.}
The CWE mapping also reveals notable gaps. Of the 61 unique CWE entries identified, the majority (42) appear in only a single curated paper, suggesting that many weakness classes are studied in isolation rather than recognized as systematic transaction security issues. Furthermore, several CWE categories that are highly relevant to modern multi-context systems, such as CWE-799 (Improper Control of Interaction Frequency), CWE-1021 (Improper Restriction of Rendered UI Layers), and CWE-441 (Unintended Proxy or Intermediary), were not identified in any of the surveyed works, despite their clear applicability to cross-context transactional scenarios involving intermediaries, rate-limited operations, and layered interfaces. This underrepresentation likely reflects the literature's heavy concentration on Generation~III DLT systems at the expense of the broader transaction security landscape, and it highlights opportunities for future research, particularly in Generation~IV systems where cross-context intermediation and interaction-frequency control are critical concerns.

\begin{tcolorbox}[title=\textbf{Insight from Security Analysis}, colback=white!95!black, colframe=gray!75!black, colbacktitle=gray!75!black, coltitle=white, boxsep=1.5pt, left=1mm, right=1mm, top=1mm, bottom=1mm]
CWE-based classification reveals that race conditions, emergent-resource exploitation, and authentication bypass are the dominant weakness classes in transaction security. Many weakness classes recur across all four evolutionary generations, validating a cross-domain analytical approach. As systems grow more complex, the threat model shifts from implementation-level bugs toward composition-level failures arising from unanticipated interactions between correctly functioning components.
\end{tcolorbox}

\section{Fundamental Properties of Modern TPS}\label{sec:rancid}

The ACID properties of Atomicity, Consistency, Isolation, and Durability have been foundational to transaction processing since their codification in the early database era~\cite{grayTransactionConceptVirtues1981, grayTransactionProcessingConcepts1994}. Together, they ensure that transactions either succeed or fail cleanly (Atomicity), that the system's invariants are preserved (Consistency), that concurrent transactions do not interfere with one another (Isolation), and that committed changes persist despite failures (Durability)~\cite{kleppmannDesigningDataintensiveApplications2017}. These properties have proven remarkably durable across generations: they remain relevant to distributed databases, DLTs, and modern multi-context systems alike. However, our cross-generational analysis reveals that ACID alone is insufficient to capture two properties that are fundamental to secure and correct transaction processing in modern systems.

In this section, we review how ACID properties manifest across generations, identify the two missing properties, and introduce RANCID as an extended framework. We then analyze the prevalence of all six properties across the surveyed literature.

\subsection{ACID Across Generations}\label{sec:rancid:acid}

Although the four ACID properties were formalized for centralized databases, each has evolved in meaning and implementation complexity as transaction processing systems have advanced.

\emph{Atomicity} in Generation~I systems is enforced through local rollback mechanisms such as write-ahead logging. In distributed databases (Generation~II), atomicity requires multi-phase commit protocols (such as two-phase commit (2PC)) that coordinate across replicas~\cite{bernsteinConcurrencyControlRecovery1987}. In DLTs (Generation~III), atomicity takes on a different character: a transaction is either included in a confirmed block or it is not, with the blockchain's consensus mechanism serving as the arbiter. In Generation~IV systems, achieving atomicity across heterogeneous contexts, each with its own commit semantics, remains an open challenge.

\emph{Consistency} is notable for being the least formally defined ACID property; Kleppmann~\cite{kleppmannDesigningDataintensiveApplications2017} observes that it was originally included primarily to complete the acronym. In practice, consistency is application-dependent and is enforced through constraints and invariants. In Generation~III systems, smart contract logic serves as the consistency mechanism, but as the DAO attack demonstrated, logical correctness of the code does not guarantee consistency with the designer's intent~\cite{shierUnderstandingRevolutionaryFlawed2017}.

\emph{Isolation} ensures that concurrent transactions do not interfere with one another. Violations of isolation underlie many of the race-condition vulnerabilities (CWE-362, CWE-367) identified as dominant in \S\ref{sec:sec:patterns}. In DLTs, transaction ordering is determined by miners or validators, creating opportunities for front-running and sandwich attacks that exploit the gap between transaction submission and confirmation~\cite{daianFlashBoys202020, zhangCombattingFrontRunningSmart2023}.

\emph{Durability} in centralized and distributed databases is enforced through persistent storage and replication. In blockchain systems, durability is probabilistic: a transaction in a given block becomes increasingly durable as subsequent blocks are appended, with heuristics such as Bitcoin's six-confirmation rule providing practical thresholds~\cite{hicksWhenCanYou2025}. Dembo et al.~\cite{demboEverythingRaceNakamoto2020} have shown that certain blocks, termed ``Nakamoto blocks,'' provably guarantee the durability of all transactions up to and including them, although such blocks cannot be identified explicitly.

\subsection{Beyond ACID: Real-Timeness and $N$-many Contexts}\label{sec:rancid:beyond}

Our analysis of security issues across all four generations reveals two properties that are increasingly critical to modern transactional systems but are not captured by ACID.

\paragraph{Real-timeness (R)}
A transaction exhibits real-time constraints when its correctness depends not only on logical completion but also on completion within a bounded time window. We distinguish two levels following standard real-time systems terminology~\cite{stankovicRealtimeTransactions1988}:

\begin{itemize}[leftmargin=*]
    \item A \emph{hard real-time constraint} is one in which failure to complete the transaction within the deadline renders the result unacceptable or catastrophic; examples include an anti-lock braking system that must engage within milliseconds, or a flash loan that must be repaid within a single block time.
    \item A \emph{soft real-time constraint} is one in which timely completion is preferred but delayed completion can be tolerated by the system; examples include fraud detection alerts on credit card transactions~\cite{quahRealtimeCreditCard2008} or IP packet retransmission.
\end{itemize}

Real-time transaction processing has been studied for decades in specialized contexts such as real-time databases~\cite{lelannDistributedSystemRealTime1981, georgeSecureTransactionProcessing1997} and real-time operating systems~\cite{eTimedrivenSchedulerRealtime1986}. However, real-time constraints have become pervasive in modern systems: autonomous vehicles require control-loop transactions to complete within milliseconds, services such as Uber continuously process petabytes of data under real-time constraints across heterogeneous subsystems~\cite{fuRealtimeDataInfrastructure2021}, and DeFi applications impose hard deadlines through block-time-bounded operations~\cite{daianFlashBoys202020, chenFlashSynFlashLoan2024}.

From a security perspective, real-time constraints create a distinctive class of vulnerabilities. An attacker who can delay a time-sensitive transaction, whether through network congestion, resource exhaustion, or deliberate interference with consensus, can cause the transaction to miss its deadline, potentially with cascading consequences. Flash loan attacks critically depend on the hard real-time constraint of single-block execution: the attacker exploits the guarantee that the entire attack sequence will either complete atomically within the block or not at all~\cite{chenFlashSynFlashLoan2024}. Front-running attacks similarly exploit timing: by observing pending transactions in the mempool and submitting competing transactions with higher gas fees, an attacker can manipulate the effective ordering and timing of others' operations~\cite{daianFlashBoys202020}.

\paragraph{$N$-many Contexts (N)}
A transaction exhibits the $N$-many contexts property when its execution requires coordination across $N \geq 2$ distinct and potentially heterogeneous operational contexts. Here, a \emph{context} refers to an independent system, subsystem, or domain with its own state, trust assumptions, failure modes, and possibly distinct administrative authority.

While Generation~II introduced multi-system transactions (inter-database replication) and Generation~III added inter-node coordination (consensus, smart contracts), these typically involve homogeneous contexts, multiple instances of the same type of system. Generation~IV systems are distinguished by the heterogeneity of their contexts: a single transaction may span software systems, physical sensors, hardware actuators, payment networks, and external data sources, each with fundamentally different operational characteristics.

Consider the contexts involved in a ride-hailing transaction: the end-user's mobile application, the driver's application, multiple payment processors, geographic tracking services, and, in the case of food delivery, one or more restaurant ordering systems, each with its own menu and backend~\cite{fuRealtimeDataInfrastructure2021}. Similarly, the flight control of an autonomous drone involves transactions across altimeters, GPS receivers, propeller controllers, collision-avoidance sensors, and communication modules, each a distinct context with its own state and failure modes.

The security implications of $N$-many contexts are substantial. Each additional context introduces new trust boundaries, communication channels, and potential points of compromise. Enforcing consistent access-control policies across heterogeneous contexts with different authorization models is a largely unsolved problem. Moreover, composition-level failures, in which individually correct components produce incorrect behavior when combined, become more likely as $N$ grows. The dominance of CWE-1229 (Creation of Emergent Resource) and CWE-841 (Improper Enforcement of Behavioral Workflow) in our classification (\S\ref{sec:sec:patterns}) directly reflects this challenge: these weaknesses arise precisely when transactions span multiple contexts in unanticipated ways.

\subsection{The RANCID Framework}\label{sec:rancid:framework}

We extend the classical ACID properties with Real-timeness~(R) and $N$-many Contexts~(N) to form the RANCID property set:

\begin{tcolorbox}[breakable, 
    enhanced jigsaw,
    colback=white!95!black, colframe=gray!60!black, boxsep=2pt, left=2mm, right=2mm, top=2mm, bottom=2mm]
\textbf{RANCID Properties:}
\begin{itemize}[leftmargin=*, topsep=2pt, itemsep=1pt]
    \item[\textbf{R}] \textbf{Real-timeness:} The system can enforce hard and/or soft timing constraints on transaction completion.
    \item[\textbf{A}] \textbf{Atomicity:} Transactions complete entirely or have no effect, with rollback across all involved contexts.
    \item[\textbf{N}] \textbf{$N$-many Contexts:} The system supports transactional coordination across $N \geq 2$ heterogeneous contexts.
    \item[\textbf{C}] \textbf{Consistency:} Transactions preserve system invariants and business-logic constraints.
    \item[\textbf{I}] \textbf{Isolation:} Concurrent transactions do not interfere with one another's execution or results.
    \item[\textbf{D}] \textbf{Durability:} Committed transaction effects persist despite subsequent failures.
\end{itemize}
\end{tcolorbox}

\subsection{RANCID in the Surveyed Literature}\label{sec:rancid:analysis}

To assess the relevance of the extended properties, we annotated each of the 41 curated papers with the RANCID properties that are affected by, or directly related to, its security focus (Table~\ref{tab:classification}). The results strongly support the inclusion of R and N.

Among the 41 curated papers, Real-timeness~(R) is relevant to 30 (73\%) and $N$-many Contexts~(N) is relevant to 35 (85\%). Both R and N apply simultaneously to 29 papers (71\%), and only 5 papers (12\%) involve neither extended property. By comparison, the classical ACID properties appear with the following frequencies: Atomicity~29 (71\%), Consistency~39 (95\%), Isolation~34 (83\%), and Durability~34 (83\%). The two new properties are thus \emph{at least as prevalent} as Atomicity in the surveyed security literature and approach the prevalence of Isolation and Durability.

These figures demonstrate that R and N are not niche concerns but rather are fundamental to the majority of transaction security research. The high co-occurrence of R and N (71\% of papers) further suggests that modern transaction security threats frequently involve both timing constraints and multi-context coordination simultaneously, precisely the combination that ACID was not designed to address. Researchers analyzing the security of modern transactional systems will benefit from considering the full RANCID property set rather than the classical ACID properties alone.

\begin{tcolorbox}[title=\textbf{Insight from Fundamental Properties Analysis}, colback=white!95!black, colframe=gray!75!black, colbacktitle=gray!75!black, coltitle=white, boxsep=1.5pt, left=1mm, right=1mm, top=1mm, bottom=1mm]
The classical ACID properties are necessary but insufficient for modern transactional systems. Real-timeness~(R) and $N$-many Contexts~(N) are relevant to 73\% and 85\% of the curated security papers, respectively, and co-occur in 71\%. The RANCID framework captures these properties alongside ACID, providing a more complete basis for analyzing the security and correctness of contemporary transaction processing systems.
\end{tcolorbox}

\begin{table*}[htbp]
\renewcommand{\arraystretch}{0.5}

\linespread{0.9}
\renewcommand{\arraystretch}{0.5}

\setlength{\tabcolsep}{4pt}
    \centering

            \caption{Classification of transaction security--related papers based on the proposed taxonomy.}
            \label{tab:classification}
            
    \begin{tabular}{P{0.7cm}:P{3.3cm}||P{3.0cm}:P{2.0cm}:P{0.25cm}:P{0.25cm}:P{0.25cm}:P{0.25cm}:P{0.25cm}:P{0.25cm}:P{2.0cm}:P{1.5cm}}

        \arrayrulecolor{black}\toprule

        \multicolumn{1}{c}{\multirow{2}{*}{\textbf{}}} &
        \multicolumn{1}{c}{\multirow{2}{*}{\textbf{}}} & 

        \multicolumn{9}{c}{\textit{Classification Criteria (Dimensions)}} \\
        \cmidrule(lr){3-12}

        \multirow{2}{*}{\textbf{Year}} & \multirow{2}{*}{\textbf{Author(s)}} &
        \multirow{2}{*}{\small Contribution$\mathrm{^\dagger}$} & {\small Associated} & \multirow{2}{*}{\small R} & \multirow{2}{*}{\small A} & \multirow{2}{*}{\small N} & \multirow{2}{*}{\small C} & \multirow{2}{*}{\small I} & \multirow{2}{*}{\small D} & {\small Security } & {\small Evolutionary }\\
        & %
        & %
        & {\small CWEs} & & & & & & & {\small Focus} & {\small Stage} \\
        \arrayrulecolor{black}\midrule

\citeyear{grayTransactionConceptVirtues1981} & \vspace{2pt}\citeauthor{grayTransactionConceptVirtues1981}~\cite{grayTransactionConceptVirtues1981}\vspace{2pt} & \vspace{2pt}\texttt{DIS} \texttt{GEN}\vspace{2pt} & n/a &  & \checkmark &  & \checkmark & \checkmark & \checkmark & {\small \texttt{Def}} & I \\
\citeyear{chaumSecurityIdentificationTransaction1985a} & \vspace{2pt}\citeauthor{chaumSecurityIdentificationTransaction1985a}~\cite{chaumSecurityIdentificationTransaction1985a}\vspace{2pt} & \vspace{2pt}\texttt{SCH} \texttt{MDL} \texttt{DIS}\vspace{2pt} & 359, 201, 212 &  & \checkmark & \checkmark & \checkmark & \checkmark & \checkmark & {\small \texttt{Def}} \texttt{Vul} & III \\
\citeyear{clarkComparisonCommercialMilitary1987} & \vspace{2pt}\citeauthor{clarkComparisonCommercialMilitary1987}~\cite{clarkComparisonCommercialMilitary1987}\vspace{2pt} & \vspace{2pt}\texttt{CST} \texttt{EVAL} \texttt{SEC} \texttt{MDL}\vspace{2pt} & 284, 778 &  & \checkmark &  & \checkmark & \checkmark & \checkmark & {\small \texttt{Def}} \texttt{Vul} & II \\
\citeyear{steinerKerberosAuthenticationService1988} & \vspace{2pt}\citeauthor{steinerKerberosAuthenticationService1988}~\cite{steinerKerberosAuthenticationService1988}\vspace{2pt} & \vspace{2pt}\texttt{MDL}\vspace{2pt} & 306, 523 &  & \checkmark &  & \checkmark & \checkmark & \checkmark & {\small \texttt{Def}} \texttt{Vul} & II \\
\citeyear{bondAttacksCryptoprocessorTransaction2001} & \vspace{2pt}\citeauthor{bondAttacksCryptoprocessorTransaction2001}~\cite{bondAttacksCryptoprocessorTransaction2001}\vspace{2pt} & \vspace{2pt}\texttt{CST} \texttt{ATT} \texttt{EVAL}\vspace{2pt} & 320, 284, 326 &  & \checkmark &  & \checkmark & \checkmark &  & \texttt{Att} {\small \texttt{Def}} & I \\
\citeyear{ammannRecoveryMaliciousTransactions2002} & \vspace{2pt}\citeauthor{ammannRecoveryMaliciousTransactions2002}~\cite{ammannRecoveryMaliciousTransactions2002}\vspace{2pt} & \vspace{2pt}\texttt{SOL} \texttt{MDL} \texttt{EVAL}\vspace{2pt} & 862 & \checkmark & \checkmark & \checkmark & \checkmark & \checkmark & \checkmark & {\small \texttt{Def}} \texttt{Vul} & II \\
\citeyear{gilbertBrewersConjectureFeasibility2002} & \vspace{2pt}\citeauthor{gilbertBrewersConjectureFeasibility2002}~\cite{gilbertBrewersConjectureFeasibility2002}\vspace{2pt} & \vspace{2pt}\texttt{TRY}\vspace{2pt} & 820, 341 &  & \checkmark &  & \checkmark & \checkmark & \checkmark & \texttt{Vul} & II \\
\citeyear{fooADEPTSAdaptiveIntrusion2005} & \vspace{2pt}\citeauthor{fooADEPTSAdaptiveIntrusion2005}~\cite{fooADEPTSAdaptiveIntrusion2005}\vspace{2pt} & \vspace{2pt}\texttt{SOL} \texttt{MDL} \texttt{EVAL}\vspace{2pt} & 787, 770 & \checkmark &  & \checkmark & \checkmark & \checkmark &  & {\small \texttt{Def}} \texttt{Att} \texttt{Vul} & II \\
\citeyear{murdochChipPINBroken2010} & \vspace{2pt}\citeauthor{murdochChipPINBroken2010}~\cite{murdochChipPINBroken2010}\vspace{2pt} & \vspace{2pt}\texttt{ATT}\vspace{2pt} & 305, 290, 669 & \checkmark & \checkmark & \checkmark & \checkmark & \checkmark & \checkmark & \texttt{Att} & II \\
\citeyear{xieIntegrityDataAttacks2011} & \vspace{2pt}\citeauthor{xieIntegrityDataAttacks2011}~\cite{xieIntegrityDataAttacks2011}\vspace{2pt} & \vspace{2pt}\texttt{ATT} \texttt{EVAL} \texttt{CST} \texttt{SUR}\vspace{2pt} & 346, 349, 290 & \checkmark &  & \checkmark & \checkmark & \checkmark &  & \texttt{Att} \texttt{Vul} {\small \texttt{Def}} & II \\
\citeyear{karameTwoBitcoinsPrice2012} & \vspace{2pt}\citeauthor{karameTwoBitcoinsPrice2012}~\cite{karameTwoBitcoinsPrice2012}\vspace{2pt} & \vspace{2pt}\texttt{ATT} \texttt{EVAL} \texttt{CST}\vspace{2pt} & 367, 346, 358 & \checkmark & \checkmark & \checkmark & \checkmark & \checkmark & \checkmark & \texttt{Att} \texttt{Vul} {\small \texttt{Def}} & III \\
\citeyear{krollEconomicsBitcoinMining2013} & \vspace{2pt}\citeauthor{krollEconomicsBitcoinMining2013}~\cite{krollEconomicsBitcoinMining2013}\vspace{2pt} & \vspace{2pt}\texttt{TRY} \texttt{DIS} \texttt{GEN}\vspace{2pt} & 367, 362, 358 & \checkmark & \checkmark & \checkmark & \checkmark & \checkmark &  & \texttt{Vul} \texttt{Att} {\small \texttt{Def}} & III \\
\citeyear{bondChipSkimCloning2014} & \vspace{2pt}\citeauthor{bondChipSkimCloning2014}~\cite{bondChipSkimCloning2014}\vspace{2pt} & \vspace{2pt}\texttt{ATT} \texttt{EVAL} \texttt{ATT}\vspace{2pt} & 340, 294, 331 & \checkmark &  & \checkmark & \checkmark &  & \checkmark & \texttt{Att} & II \\
\citeyear{deckerBitcoinTransactionMalleability2014} & \vspace{2pt}\citeauthor{deckerBitcoinTransactionMalleability2014}~\cite{deckerBitcoinTransactionMalleability2014}\vspace{2pt} & \vspace{2pt}\texttt{EMP} \texttt{CST} \texttt{DIS}\vspace{2pt} & 353, 347, 1287 & \checkmark & \checkmark & \checkmark & \checkmark & \checkmark & \checkmark & \texttt{Vul} \texttt{Att} {\small \texttt{Def}} & III \\
\citeyear{bonneauWhyBuyWhen2016} & \vspace{2pt}\citeauthor{bonneauWhyBuyWhen2016}~\cite{bonneauWhyBuyWhen2016}\vspace{2pt} & \vspace{2pt}\texttt{ATT} \texttt{DIS}\vspace{2pt} & 1229, 514, 362 & \checkmark & \checkmark & \checkmark & \checkmark & \checkmark & \checkmark & \texttt{Att} \texttt{Vul} & III \\
\citeyear{nayakStubbornMiningGeneralizing2016} & \vspace{2pt}\citeauthor{nayakStubbornMiningGeneralizing2016}~\cite{nayakStubbornMiningGeneralizing2016}\vspace{2pt} & \vspace{2pt}\texttt{ATT} \texttt{DIS}\vspace{2pt} & 1229, 221, 514 & \checkmark & \checkmark & \checkmark & \checkmark & \checkmark & \checkmark & \texttt{Att} & III \\
\citeyear{apostolakiHijackingBitcoinRouting2017} & \vspace{2pt}\citeauthor{apostolakiHijackingBitcoinRouting2017}~\cite{apostolakiHijackingBitcoinRouting2017}\vspace{2pt} & \vspace{2pt}\texttt{ATT} \texttt{EMP}\vspace{2pt} & 319, 924, 290 & \checkmark &  & \checkmark & \checkmark &  & \checkmark & \texttt{Att} \texttt{Vul} & III \\
\citeyear{atzeiSurveyAttacksEthereum2017} & \vspace{2pt}\citeauthor{atzeiSurveyAttacksEthereum2017}~\cite{atzeiSurveyAttacksEthereum2017}\vspace{2pt} & \vspace{2pt}\texttt{SUR} \texttt{GEN} \texttt{CST} \texttt{SEC}\vspace{2pt} & 252, 367, 338 & \checkmark & \checkmark & \checkmark & \checkmark & \checkmark &  & \texttt{Vul} \texttt{Att} & III \\
\citeyear{kumarTraceabilityAnalysisMoneros2017} & \vspace{2pt}\citeauthor{kumarTraceabilityAnalysisMoneros2017}~\cite{kumarTraceabilityAnalysisMoneros2017}\vspace{2pt} & \vspace{2pt}\texttt{ATT} \texttt{EVAL}\vspace{2pt} & 573, 205, 1229 &  & \checkmark & \checkmark & \checkmark &  & \checkmark & \texttt{Att} {\small \texttt{Def}} & III \\
\citeyear{shierUnderstandingRevolutionaryFlawed2017} & \vspace{2pt}\citeauthor{shierUnderstandingRevolutionaryFlawed2017}~\cite{shierUnderstandingRevolutionaryFlawed2017}\vspace{2pt} & \vspace{2pt}\texttt{CST} \texttt{DIS}\vspace{2pt} & 841, 283, 367 & \checkmark & \checkmark & \checkmark & \checkmark & \checkmark & \checkmark & \texttt{Att} \texttt{Vul} & III \\
\citeyear{eyalMajorityNotEnough2018} & \vspace{2pt}\citeauthor{eyalMajorityNotEnough2018}~\cite{eyalMajorityNotEnough2018}\vspace{2pt} & \vspace{2pt}\texttt{ATT} \texttt{SCH}\vspace{2pt} & 1229, 221, 362 &  & \checkmark & \checkmark & \checkmark & \checkmark & \checkmark & \texttt{Att} & III \\
\citeyear{gaziStakeBleedingAttacksProofofStake2018} & \vspace{2pt}\citeauthor{gaziStakeBleedingAttacksProofofStake2018}~\cite{gaziStakeBleedingAttacksProofofStake2018}\vspace{2pt} & \vspace{2pt}\texttt{ATT} \texttt{SOL} \texttt{TRY}\vspace{2pt} & 362, 294 & \checkmark &  & \checkmark & \checkmark &  & \checkmark & \texttt{Att} \texttt{Vul} {\small \texttt{Def}} & III \\
\citeyear{kruppTeEtherGnawingEthereum2018} & \vspace{2pt}\citeauthor{kruppTeEtherGnawingEthereum2018}~\cite{kruppTeEtherGnawingEthereum2018}\vspace{2pt} & \vspace{2pt}\texttt{SOL} \texttt{ATT} \texttt{MDL} \texttt{EVAL} \texttt{CST}\vspace{2pt} & 205, 1229, 94 &  &  & \checkmark & \checkmark &  & \checkmark & \texttt{Att} \texttt{Vul} & III \\
\citeyear{moserEmpiricalAnalysisTraceability2018} & \vspace{2pt}\citeauthor{moserEmpiricalAnalysisTraceability2018}~\cite{moserEmpiricalAnalysisTraceability2018}\vspace{2pt} & \vspace{2pt}\texttt{EMP} \texttt{HEU} \texttt{SCH} \texttt{EVAL}\vspace{2pt} & 359, 208, 334 &  &  & \checkmark & \checkmark & \checkmark & \checkmark & \texttt{Vul} {\small \texttt{Def}} & III \\
\citeyear{rodlerSereumProtectingExisting2018} & \vspace{2pt}\citeauthor{rodlerSereumProtectingExisting2018}~\cite{rodlerSereumProtectingExisting2018}\vspace{2pt} & \vspace{2pt}\texttt{SOL} \texttt{EVAL} \texttt{ATT}\vspace{2pt} & 367, 667 & \checkmark & \checkmark & \checkmark & \checkmark & \checkmark & \checkmark & {\small \texttt{Def}} \texttt{Att} & III \\
\citeyear{tsankovSecurifyPracticalSecurity2018} & \vspace{2pt}\citeauthor{tsankovSecurifyPracticalSecurity2018}~\cite{tsankovSecurifyPracticalSecurity2018}\vspace{2pt} & \vspace{2pt}\texttt{SOL}\vspace{2pt} & 840, 708, 471 & \checkmark & \checkmark & \checkmark & \checkmark & \checkmark & \checkmark & {\small \texttt{Def}} \texttt{Vul} & III \\
\citeyear{athalyeNotaryDeviceSecure2019} & \vspace{2pt}\citeauthor{athalyeNotaryDeviceSecure2019}~\cite{athalyeNotaryDeviceSecure2019}\vspace{2pt} & \vspace{2pt}\texttt{SOL} \texttt{EVAL}\vspace{2pt} & 125, 497 &  &  & \checkmark &  & \checkmark &  & {\small \texttt{Def}} \texttt{Vul} & II \\
\citeyear{brentEthainterSmartContract2020} & \vspace{2pt}\citeauthor{brentEthainterSmartContract2020}~\cite{brentEthainterSmartContract2020}\vspace{2pt} & \vspace{2pt}\texttt{SOL} \texttt{ATT} \texttt{MDL} \texttt{SEC} \texttt{EVAL}\vspace{2pt} & 1039, 841, 749 & \checkmark & \checkmark & \checkmark & \checkmark & \checkmark & \checkmark & \texttt{Vul} \texttt{Att} & III \\
\citeyear{daianFlashBoys202020} & \vspace{2pt}\citeauthor{daianFlashBoys202020}~\cite{daianFlashBoys202020}\vspace{2pt} & \vspace{2pt}\texttt{EMP} \texttt{SEC} \texttt{MDL} \texttt{SOL}\vspace{2pt} & 1229, 362, 385 & \checkmark & \checkmark & \checkmark & \checkmark & \checkmark & \checkmark & \texttt{Vul} & IV \\
\citeyear{harrisFloodLootSystemic2020} & \vspace{2pt}\citeauthor{harrisFloodLootSystemic2020}~\cite{harrisFloodLootSystemic2020}\vspace{2pt} & \vspace{2pt}\texttt{ATT} \texttt{EVAL}\vspace{2pt} & 1229, 385, 362 & \checkmark & \checkmark & \checkmark & \checkmark & \checkmark & \checkmark & \texttt{Att} {\small \texttt{Def}} & III \\
\citeyear{tramerRemoteSideChannelAttacks2020} & \vspace{2pt}\citeauthor{tramerRemoteSideChannelAttacks2020}~\cite{tramerRemoteSideChannelAttacks2020}\vspace{2pt} & \vspace{2pt}\texttt{ATT}\vspace{2pt} & 385, 208, 1229 & \checkmark &  & \checkmark & \checkmark &  & \checkmark & \texttt{Att} & III \\
\citeyear{basinEMVStandardBreak2021} & \vspace{2pt}\citeauthor{basinEMVStandardBreak2021}~\cite{basinEMVStandardBreak2021}\vspace{2pt} & \vspace{2pt}\texttt{MDL} \texttt{SEC} \texttt{ATT} \texttt{SOL}\vspace{2pt} & 1390, 294, 290 & \checkmark &  & \checkmark & \checkmark &  & \checkmark & \texttt{Att} {\small \texttt{Def}} & II \\
\citeyear{liMimosaProtectingPrivate2021} & \vspace{2pt}\citeauthor{liMimosaProtectingPrivate2021}~\cite{liMimosaProtectingPrivate2021}\vspace{2pt} & \vspace{2pt}\texttt{SOL} \texttt{EVAL} \texttt{EMP}\vspace{2pt} & 316, 125, 226 & \checkmark & \checkmark &  &  & \checkmark & \checkmark & {\small \texttt{Def}} \texttt{Vul} \texttt{Att} & II \\
\citeyear{tsabaryMADHTLCBecauseHTLC2021} & \vspace{2pt}\citeauthor{tsabaryMADHTLCBecauseHTLC2021}~\cite{tsabaryMADHTLCBecauseHTLC2021}\vspace{2pt} & \vspace{2pt}\texttt{MDL} \texttt{SOL} \texttt{SEC} \texttt{EVAL}\vspace{2pt} & 841, 362, 673 & \checkmark & \checkmark & \checkmark & \checkmark & \checkmark & \checkmark & {\small \texttt{Def}} \texttt{Att} & III \\
\citeyear{zhangCombattingFrontRunningSmart2023} & \vspace{2pt}\citeauthor{zhangCombattingFrontRunningSmart2023}~\cite{zhangCombattingFrontRunningSmart2023}\vspace{2pt} & \vspace{2pt}\texttt{EMP} \texttt{MDL} \texttt{SCH} \texttt{BMK}\vspace{2pt} & 362, 270, 402 & \checkmark &  & \checkmark & \checkmark & \checkmark & \checkmark & \texttt{Att} & III \\
\citeyear{chenFlashSynFlashLoan2024} & \vspace{2pt}\citeauthor{chenFlashSynFlashLoan2024}~\cite{chenFlashSynFlashLoan2024}\vspace{2pt} & \vspace{2pt}\texttt{SOL} \texttt{SCH} \texttt{ATT} \texttt{EVAL}\vspace{2pt} & 1039 & \checkmark & \checkmark & \checkmark & \checkmark & \checkmark & \checkmark & \texttt{Att} & IV \\
\citeyear{garayBitcoinBackboneProtocol2024} & \vspace{2pt}\citeauthor{garayBitcoinBackboneProtocol2024}~\cite{garayBitcoinBackboneProtocol2024}\vspace{2pt} & \vspace{2pt}\texttt{TRY} \texttt{SEC} \texttt{DIS}\vspace{2pt} & 684, 290, 1240 & \checkmark & \checkmark & \checkmark & \checkmark & \checkmark & \checkmark & \texttt{Vul} & III \\
\citeyear{markantonakisUsingAmbientSensors2024} & \vspace{2pt}\citeauthor{markantonakisUsingAmbientSensors2024}~\cite{markantonakisUsingAmbientSensors2024}\vspace{2pt} & \vspace{2pt}\texttt{EMP} \texttt{EVAL} \texttt{SCH}\vspace{2pt} & 294, 1390, 346 & \checkmark & \checkmark & \checkmark & \checkmark & \checkmark & \checkmark & \texttt{Vul} & IV \\
\citeyear{wuDeFiRangerDetectingDeFi2024} & \vspace{2pt}\citeauthor{wuDeFiRangerDetectingDeFi2024}~\cite{wuDeFiRangerDetectingDeFi2024}\vspace{2pt} & \vspace{2pt}\texttt{SOL} \texttt{MDL}\vspace{2pt} & 1039, 1229, 841 & \checkmark &  & \checkmark & \checkmark & \checkmark &  & {\small \texttt{Def}} \texttt{Vul} & IV \\
\citeyear{wuSafeguardingBlockchainEcosystem2025} & \vspace{2pt}\citeauthor{wuSafeguardingBlockchainEcosystem2025}~\cite{wuSafeguardingBlockchainEcosystem2025}\vspace{2pt} & \vspace{2pt}\texttt{MDL} \texttt{EMP} \texttt{EVAL}\vspace{2pt} & 841, 347, 940 & \checkmark & \checkmark & \checkmark & \checkmark & \checkmark & \checkmark & {\small \texttt{Def}} \texttt{Vul} \texttt{Att} & III \\
\citeyear{zhangManipulatedTransactionCollision2025} & \vspace{2pt}\citeauthor{zhangManipulatedTransactionCollision2025}~\cite{zhangManipulatedTransactionCollision2025}\vspace{2pt} & \vspace{2pt}\texttt{ATT} \texttt{EMP} \texttt{EVAL} \texttt{SOL}\vspace{2pt} & 367, 362, 400 & \checkmark & \checkmark & \checkmark & \checkmark & \checkmark & \checkmark & \texttt{Att} {\small \texttt{Def}} \texttt{Vul} & III \\

\hline \arrayrulecolor{black}\bottomrule \multicolumn{12}{c}{\small $\mathrm{^\dagger}$ \texttt{BK} --- Book; \texttt{CST} --- Case Study; \texttt{DIS} --- Discussion; \texttt{EMP} --- Empirical Study; \texttt{EVAL} --- Evaluation; \texttt{GEN} --- General Study;} \\

\multicolumn{12}{c}{\small  \texttt{HEU} --- Heuristics; \texttt{BMK} --- Proposal of Novel Benchmark; \texttt{MDL} --- Proposal of Novel Model; \texttt{SCH} --- Proposal of Novel Scheme;} \\ 

\multicolumn{12}{c}{\small  \texttt{SOL} --- Proposal of Novel Solution; \texttt{TRY} --- Proposal of Novel Theory; \texttt{SEC} --- Security Analysis; \texttt{SUR} --- Survey. } \\ \arrayrulecolor{black}\hline \end{tabular} \end{table*}

\section{Curated Survey}\label{sec:classification}

The field of transaction processing security spans five decades and encompasses a large and diverse body of research. Researchers entering this field may be overwhelmed by its sheer breadth, and the heavy concentration of work on Generation~III systems (Figure~\ref{fig:gens_bar}) can obscure foundational issues that cut across generations. The goal of our curated survey is to distill a representative, high-quality subset of the literature that provides a structured entry point for researchers studying the evolution, security, and fundamental properties of transactional systems.

\subsection{Search and Selection Methodology}\label{sec:classification:method}

\begin{figure}
    \centering
    \includegraphics[width=0.95\linewidth]{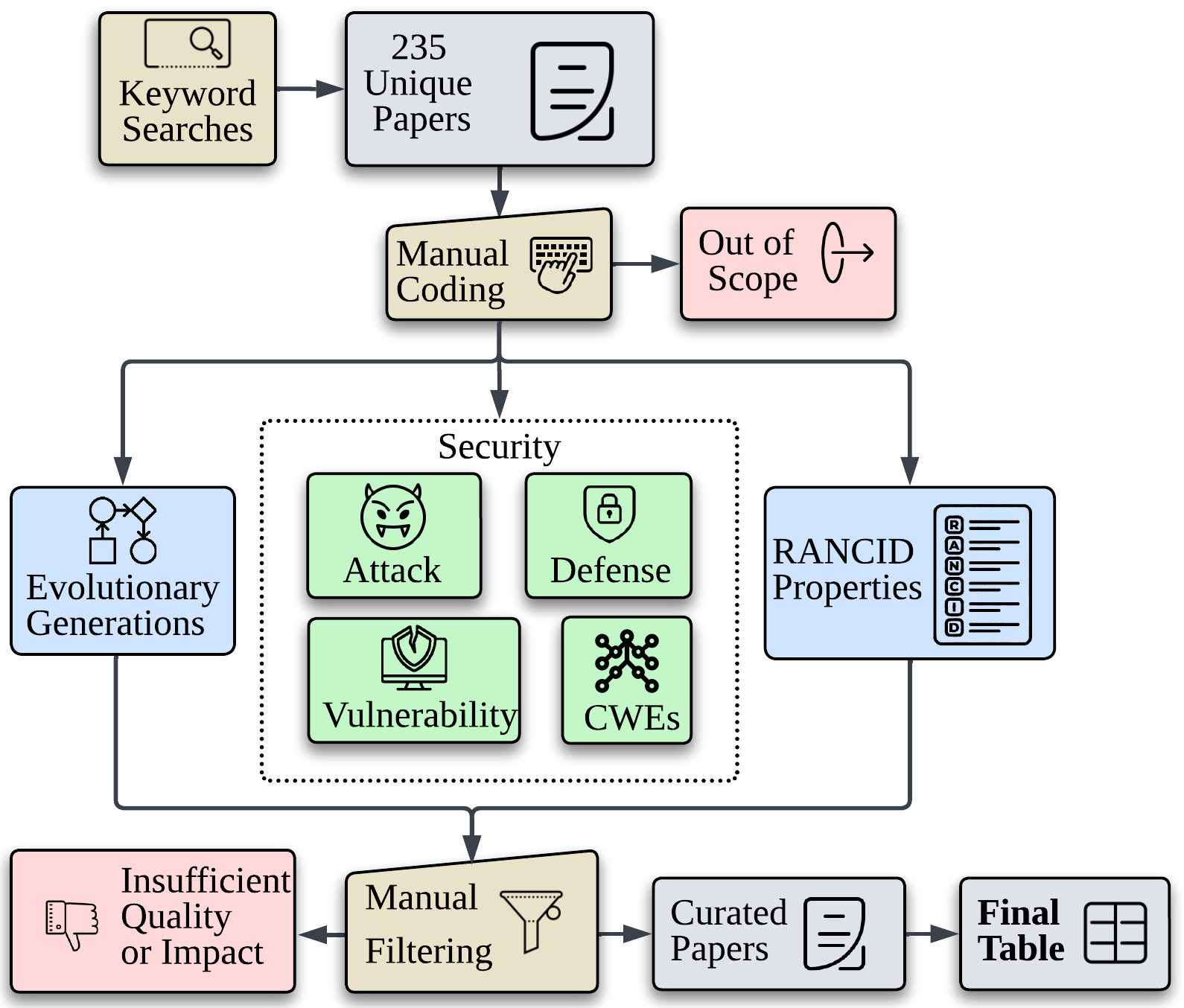}
    \caption{Process flow for identifying and classifying papers. From an initial pool of 235 papers, 163 were retained after scope filtering and classified along five dimensions; 41 were selected for the curated survey.}
    \label{fig:process_flow}
\end{figure}

Our search and selection process, illustrated in Figure~\ref{fig:process_flow}, proceeded in three stages.

\paragraph{Stage 1: Identification.}
We searched Google Scholar using keyword phrases including ``transaction security,'' ``transaction attack,'' ``transaction defense,'' ``security of transaction processing,'' and variations. This initial search yielded approximately 120 candidate papers. We then performed backward snowballing, examining the references of each candidate, to identify additional relevant works, ultimately assembling a pool of 235 unique papers.

\paragraph{Stage 2: Scope filtering.}
We reviewed all 235 papers and excluded 72 that fell outside the scope of our survey. Exclusion criteria included tangential relevance to transaction security (e.g., papers on general network security with only incidental mention of transactions), insufficient quality or rigor for inclusion in an SoK, and inaccessibility (e.g., papers behind paywalls with no preprint available). This stage retained 163 in-scope papers. Each was classified along five dimensions: evolutionary generation, contribution type (e.g., case study, novel model, empirical study), primary security focus (attack, defense, or vulnerability), associated CWE entries, and affected RANCID properties.

\paragraph{Stage 3: Quality curation.}
From the 163 in-scope papers, we selected a curated subset intended to be both high in quality and broad in coverage. Selection criteria for the curated set included: (i)~seminal or foundational status within a subfield, (ii)~high citation impact or publication at a top-tier venue, (iii)~novelty of contribution (e.g., first demonstration of a new attack class or defense technique), and (iv)~representativeness across evolutionary generations, security focus categories, and time periods. We aimed to ensure that the curated set would not merely reflect the literature's concentration on Generation~III systems but would provide balanced coverage of all four generations.

\subsection{Curation Rationale}\label{sec:classification:rationale}

The curated set of 41 papers was selected to span the full breadth of transaction security research. For Generation~I and~II systems, we included foundational works such as Gray's~\cite{grayTransactionConceptVirtues1981} formalization of the ACID properties, Clark and Wilson's~\cite{clarkComparisonCommercialMilitary1987} integrity model for commercial transaction systems, and Steiner et al.'s~\cite{steinerKerberosAuthenticationService1988} Kerberos authentication protocol, all of which remain influential in modern access-control and authorization research. Payment transaction security is represented by multiple works on the EMV standard~\cite{murdochChipPINBroken2010, bondChipSkimCloning2014, basinEMVStandardBreak2021}, and hardware-level transaction security is covered by works on cryptoprocessor attacks~\cite{bondAttacksCryptoprocessorTransaction2001} and secure transaction hardware~\cite{athalyeNotaryDeviceSecure2019}.

For the large body of Generation~III research, we prioritized works that either established fundamental results or opened new lines of inquiry. These include Garay et al.'s~\cite{garayBitcoinBackboneProtocol2024} first rigorous proof of the Bitcoin backbone protocol, Karame et al.'s~\cite{karameTwoBitcoinsPrice2012} seminal analysis of double-spending in fast-payment scenarios, Chaum's~\cite{chaumSecurityIdentificationTransaction1985a} early work on decentralized identity that prefigured DLTs, and the Shier et al.~\cite{shierUnderstandingRevolutionaryFlawed2017} post-mortem on the DAO attack. We also included representative works across the spectrum of smart contract attacks and defenses~\cite{atzeiSurveyAttacksEthereum2017, nayakStubbornMiningGeneralizing2016, bonneauWhyBuyWhen2016, kruppTeEtherGnawingEthereum2018, rodlerSereumProtectingExisting2018, tsankovSecurifyPracticalSecurity2018} and privacy analyses~\cite{kumarTraceabilityAnalysisMoneros2017, moserEmpiricalAnalysisTraceability2018}. For Generation~IV, we included works on flash loan attacks~\cite{daianFlashBoys202020, chenFlashSynFlashLoan2024}, cross-platform payment security~\cite{markantonakisUsingAmbientSensors2024}, and DeFi threat detection~\cite{wuDeFiRangerDetectingDeFi2024}.

\subsection{Composition and Coverage}\label{sec:classification:composition}

The full classification of the 41 curated papers appears in Table~\ref{tab:classification}. Analysis of the curated set's composition reveals several notable patterns.

\paragraph{Generational distribution.}
Among the curated papers, 2 address Generation~I, 11 address Generation~II, 24 address Generation~III, and 4 address Generation~IV. While Generation~III remains the largest category, the curated set achieves substantially broader generational coverage than the overall literature: Generation~III accounts for 59\% of curated papers versus 66\% of the full 163-paper corpus, and earlier generations are proportionally better represented.

\begin{figure}
            \centering
            \includegraphics[width=0.95\linewidth]{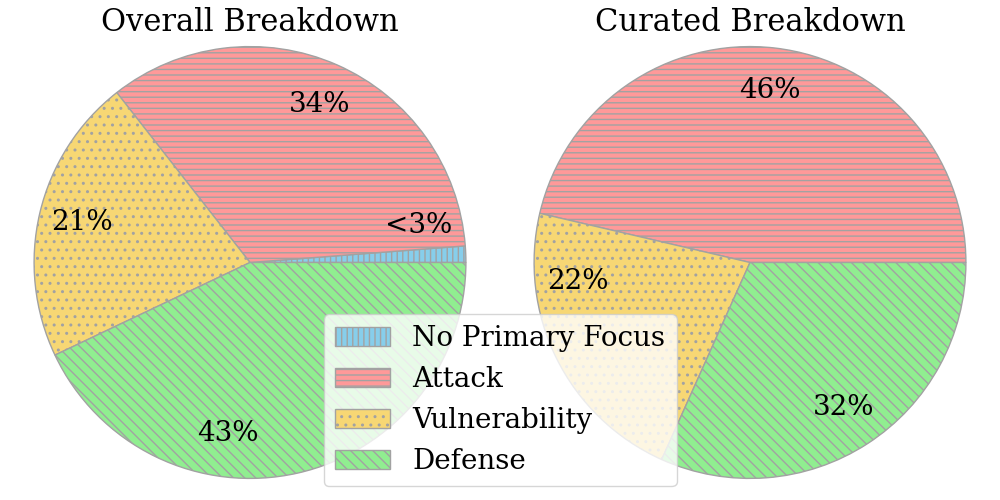}
            \caption{Primary security focus of the 163 in-scope papers (left) and the 41 curated papers (right). The curated set shifts toward attack-focused works, reflecting the higher novelty threshold of demonstrated exploits.}
            \label{fig:breakdown}
        \end{figure}

\paragraph{Security focus.}
Among the 163 in-scope papers, 56 (34\%) were primarily focused on attacks, 70 (43\%) on defense, and 35 (21\%) on vulnerabilities, with 2 being general works without a specific security focus. After curation, the 41 selected papers comprise 19 (46\%) primarily focused on attacks, 13 (32\%) on defense, and 9 (22\%) on vulnerabilities (Figure~\ref{fig:breakdown}). The shift toward attack-focused papers in the curated set reflects the generally higher novelty and impact of works that identify and demonstrate new threat classes compared to incremental defensive contributions.

\paragraph{Temporal distribution.}
The curated papers span 1981--2025, with a median publication year of 2018. Eight papers (20\%) predate 2010, establishing the historical foundations, while 25 (61\%) were published in 2017 or later, reflecting the surge of activity driven by DLT security research. The temporal spread confirms that our curation captures both the foundational era and the current research frontier.

\paragraph{Contribution types.}
The most common contribution types in the curated set are evaluations (20 papers), attack demonstrations (20), novel solutions (15), and novel models (13). Empirical studies appear in 9 papers, and theoretical contributions in 4. The prevalence of evaluation and attack-demonstration papers reflects the field's empirical orientation and the practical nature of transaction security research.

\subsection{Limitations}\label{sec:classification:limitations}

Several limitations of our methodology should be acknowledged. First, our initial search was conducted exclusively through Google Scholar, and while backward snowballing expanded the candidate pool substantially, forward citation tracking and searches of additional databases (e.g., IEEE Xplore, ACM Digital Library, Scopus) might have surfaced relevant works that we missed. Second, the curation process, while guided by explicit criteria, inevitably involves subjective judgments about quality and representativeness; a different set of curators might have made different selections, particularly among the many competitive Generation~III papers. Third, our generational classification treats each paper as belonging to a single generation, but some works, particularly those on flash loans and cross-chain protocols, straddle the boundary between Generations~III and~IV. Finally, CWE assignment required interpretive judgment, since the mapping from a paper's security focus to specific CWE entries is not always unambiguous.

\section{Conclusion and Future Work}\label{sec:conclusion}

Transaction processing systems have evolved over five decades through four major generations, from centralized databases, through distributed databases and distributed ledger technologies, to modern multi-context systems that coordinate transactions across heterogeneous cyber-physical environments with real-time constraints. In this Systematization of Knowledge, we surveyed 235 papers on transaction security, classified 163 in-scope works, and distilled a curated set of 41 high-impact or seminal papers spanning all four generations. Our analysis yields four principal findings.

First, the evolutionary taxonomy reveals a pronounced imbalance in the literature: Generation~III (DLT) systems account for 66\% of the surveyed security research, while Generation~IV multi-context systems, which represent the current and future frontier of transaction processing, remain severely underexplored. This imbalance leaves critical security questions about cross-context coordination, heterogeneous trust boundaries, and real-time transactional integrity largely unaddressed.

Second, CWE-based classification exposes structural commonalities that domain-specific analyses obscure. Race conditions and timing weaknesses (CWE-362, CWE-367) and emergent-resource exploitation (CWE-1229) are the dominant weakness classes, each appearing in 9 of the 41 curated papers. Crucially, many weakness classes recur across all four generations, manifesting as concurrency-control exploits in Generation~I, partition-based attacks in Generation~II, front-running and double-spending in Generation~III, and cross-context timing attacks in Generation~IV, further validating a unified, cross-domain approach to transaction security.

Third, the classical ACID properties, while still necessary, are insufficient for modern transactional systems. The RANCID framework extends ACID with Real-timeness~(R) and $N$-many Contexts~(N), and our analysis shows that these extended properties are not niche concerns: R is relevant to 73\% and N to 85\% of the curated papers, with both co-occurring in 71\%. Real-timeness is as prevalent as Atomicity in the surveyed security literature, underscoring that timing constraints are now a first-class concern in transaction security.

Fourth, the CWE mapping reveals a shift in the dominant threat model as systems grow more complex. In earlier generations, the primary threats are implementation-level bugs and protocol flaws. In Generations~III and~IV, the most prevalent weaknesses, CWE-1229 (Creation of Emergent Resource) and CWE-841 (Improper Enforcement of Behavioral Workflow, arise from unanticipated interactions between correctly functioning components. This shift toward composition-level failures has profound implications for how transaction security should be analyzed and tested.

\paragraph{Future work.}
Our systematization points to several concrete directions for future research. The most pressing need is for security analyses of Generation~IV systems that explicitly account for $N$-many contexts and real-time constraints, properties that are pervasive in practice but rarely treated as first-class concerns in the security literature. The dominance of composition-level weaknesses in later generations suggests a need for new analytical tools that can reason about emergent behaviors arising from the interaction of multiple transactional subsystems, potentially drawing on techniques from compositional verification and assume-guarantee reasoning. The CWE gaps identified in \S\ref{sec:sec:patterns}, including weakness classes related to interaction frequency control, unintended intermediaries, and cross-context policy enforcement, represent specific underexplored areas where targeted research could yield high impact. Finally, the RANCID framework itself invites formalization: while R and N capture important properties at a conceptual level, developing formal specifications and verification techniques for these properties across heterogeneous transactional contexts remains an open challenge. We hope that this systematization provides a foundation for such efforts and a structured entry point for researchers working toward secure transaction processing in the modern era.

\section{Acknowledgement}
This material is based upon work supported by the National Science Foundation under Grant No. 2348344. Any opinions, findings, and conclusions or recommendations expressed in this material are those of the author(s) and do not necessarily reflect the views of the National Science Foundation.

\bibliographystyle{ACM-Reference-Format}
\bibliography{cleaned}

\end{document}